\documentclass[
 twocolumn,
 amsmath, amssymb,
 aps,
 nofootinbib
]{revtex4-2}

\usepackage{natbib}
\usepackage{aas_macros}
\usepackage{booktabs} 
\usepackage{graphicx}
\usepackage{dcolumn}
\usepackage{bm}
\usepackage{hyperref}
\usepackage{orcidlink}

\usepackage[dvipsnames]{xcolor}
\usepackage[acronyms,automake,nomain]{glossaries-extra}

\newcommand{\forb}{f_{\rm orb}}
\newcommand{\forbE}{f_{\rm orb}^{(\mathrm{E})}}
\newcommand{\forbP}[1]{\forb^{(\mathrm{P}, \: #1)}}
\newcommand{\fgw}{f_{\rm GW}}
\newcommand{\fth}{f_{\rm th}}

\newcommand{\forbISCO}{\forb^{\rm (ISCO)}}

\newcommand{\msun}{M_\odot}

\newcommand{\Mc}{\mathcal{M}_{\rm c}}
\newcommand{\Mcr}{\mathcal{M}_{\rm c, r}}
\newcommand{\Mcstar}{\mathcal{M}_{\mathrm{c}, \star}}
\newcommand{\dL}{d_{\rm L}}
\newcommand{\Vc}{V_{\rm c}}
\newcommand{\dd}{\mathrm{d}}
\newcommand{\tauc}{\tau_{\rm c}}
\newcommand{\taucmin}{\tau_{\rm c}^{\rm (min)}}
\newcommand{\taucr}{\tau_{\rm c, r}}

\newcommand{\Tc}{T_{\rm c}}
\newcommand{\tc}{t_{\rm c}}

\newcommand{\Ta}{T_a}
\newcommand{\Tmax}{T^{\rm (max)}}

\newcommand{\xiI}{\vec{\xi}_{\rm I}}

\newcommand{\Tobs}{T_{\rm obs}}

\newcommand{\Nz}{N_{\rm z}}
\newcommand{\Nzbar}{\langle N_{\rm z} \rangle}

\newcommand{\Npsr}{N_{\rm psr}}
\newcommand{\SNR}{\mathrm{SNR}}
\newcommand{\SNRthresh}{\mathrm{SNR}_{\rm th}}

\newcommand{\RGW}{R_{\rm GW}}

\setabbreviationstyle[acronym]{long-short}

\newacronym[plural={Pulsar Timing Arrays},
   shortplural={PTAs}]{pta}{PTA}{Pulsar Timing Array}

\newacronym[plural={Gravitational Waves},
   shortplural={GWs}]{gw}{GW}{Gravitational Wave}

\newacronym[
    longplural={Supermassive Black Hole Binaries},
    shortplural={SMBHBs}
]{smbhb}{SMBHB}{Supermassive Black Hole Binary}

\newacronym{isco}{ISCO}{Innermost Stable Circular Orbit}

\newacronym{snr}{SNR}{Signal-to-Noise Ratio}
\newacronym{pn}{PN}{Post-Newtonian}

\makeglossaries

\begin{document}

\title{Probing past mergers of supermassive black holes with pulsar timing arrays: \\The role of pulsar terms}

\author{Hippolyte Quelquejay Leclere \orcidlink{0000-0002-6766-2004}}
\email{hippolyte.quelquejayleclere@unimib.it}
\affiliation{Dipartimento di Fisica “G. Occhialini”,
Università degli Studi di Milano-Bicocca,\\
Piazza della Scienza 3, I-20126 Milano, Italy.}

\date{\today}

\begin{abstract}
By monitoring the times of arrival of radio pulses from millisecond pulsars, Pulsar Timing Arrays (PTAs) serve as unique gravitational wave (GW) laboratories in the nanohertz band.
To date, the primary astrophysical sources of GWs targeted in this frequency range have been inspiraling supermassive black hole binaries (SMBHBs) on circular and eccentric orbits.
In this work, we demonstrate that, thanks to the so-called pulsar term in the timing residual waveform of GW signals, PTAs can probe individual SMBHBs that merged before timing observations began.
We refer to the latter as \emph{zombie binaries}.
Using SMBHB population models consistent with current PTA constraints, we find that while the probability of detecting such systems in existing PTA datasets remains low, the Square Kilometer Array observatory is expected to achieve sufficient sensitivity to have a few zombie binaries with {optimal matched filter signal-to-noise ratios} exceeding 3 in its data.
Although their confident identification might be challenging, this new class of PTA sources opens a novel window for studying the most massive SMBHBs in our local universe.
\end{abstract}

\maketitle

\section{\label{sec:intro}Introduction}

Recently, \gls*{pta} collaborations have reported compelling evidence for the presence of \gls*{gw} signals in their datasets \cite{2023A&A...678A..50E,2023ApJ...951L...8A,2023RAA....23g5024X,2023ApJ...951L...6R,2025MNRAS.536.1489M}.
The primary source of these nanohertz \glspl*{gw} is believed to be the population of inspiraling \glspl*{smbhb} \cite{1995ApJ...446..543R,2003ApJ...583..616J}, which are expected to form within the hierarchical model of galaxy formation \cite{1977ApJ...217L.125O,1993MNRAS.262..627L}.
If supermassive black hole pairs manage to efficiently reach sub-parsec separation within a Hubble time \cite{2003ApJ...596..860M} -- through dynamical friction and stellar hardening \cite{1980Natur.287..307B,1996NewA....1...35Q} -- they become efficient sources of \glspl*{gw} in the $1-100$ nanohertz band, until they merge at orbital frequencies of hundreds of nanohertz.
Therefore, binaries emitting in the \gls*{pta} band are generally thousands to millions of years from merger and are thus expected to be observed as quasi-monochromatic sources over \glspl*{pta} timescale.

{Up to now, \glspl*{pta} have searched for \glspl*{smbhb} using three main approaches.}
The first involves detecting the stochastic \gls*{gw} background produced by the population of binaries, which is expected to imprint a quadrupolar correlation pattern among pulsars' timing residuals, known as the Hellings-Downs curve \cite{1983ApJ...265L..39H}.
The presence of such a signal has been reported with strong evidence by various \glspl*{pta}, along with a characterization of its power spectral density in the $1-10$ nanohertz range, mostly consistent across collaborations \cite{2024ApJ...966..105A}.
Although potentially degenerate with, for example, other cosmological \gls*{gw} backgrounds \cite{2023ApJ...951L..11A,2024A&A...685A..94E}, the identification of significant anisotropies \cite{2013PhRvD..88f2005M,2024PhRvD.110f3526J}, non-stationarities \cite{2025PhRvD.111b3047F} or non-Gaussianities \cite{2025arXiv251109659L,2026PhRvD.113d3047F} could help proving the astrophysical origin of the stochastic signal.
But collaborations have not yet found strong evidence of such features \cite{2024PhRvD.109l3010F,2025MNRAS.536.1501G,2023ApJ...956L...3A,2026PhRvD.113d3042C}.

The second, more direct approach is the search for \gls*{gw} signals from individual binaries, inspiraling on circular or eccentric orbits, in the \gls*{pta} data \cite{2001ApJ...562..297L,2009MNRAS.394.2255S,2016ApJ...817...70T}.
The pulsars' timing residuals induced by an inspiraling \gls*{smbhb} are made of two terms, respectively produced by the metric perturbation at the pulse emissions (pulsar term) and their reception (Earth term).
{Assuming that the \gls*{smbhb} dynamics is driven by \glspl*{gw} emission, a deterministic template for both the Earth and pulsar terms in each pulsar timing residuals can be derived and searched for in \gls*{pta} data.}
So far, no conclusive evidence for the presence of such a system has been reported \cite{2023ApJ...951L..50A,2024A&A...690A.118E}.

{Finally, \glspl*{pta} are also sensitive to the \gls*{gw} memory signal produced by \gls*{smbhb} coalescences. 
If the memory wavefront from a binary merger passes through the Earth during the detector's observation period, it will -- at leading order -- produce deterministic linear trends in the pulsars' timing residuals (see \cite{2010ApJ...718.1400F,2010MNRAS.401.2372V,2014ApJ...788..141M,2012ApJ...752...54C} for further discussion of memory signals in \gls*{pta} data).
While in principle observable, these events are expected to be rare, as they require a sufficiently massive and nearby \gls*{smbhb} to merge during the \gls*{pta} observing campaign.
Nonetheless, searches for memory burst signatures in \gls*{pta} data have been carried out and have yielded constraints on the merger rate of \glspl*{smbhb} in our local universe \cite{2015MNRAS.446.1657W,2026ApJ...996L...9T}.}

Pitkin \cite{Pitkin_2008} was the first to notice that the signal from an individual \gls*{smbhb} could take a distinct form in the case of a \gls*{smbhb} merging in the millihertz band of the Laser Interferometer Space Antenna (LISA) detector.
In this scenario, {the Earth terms associated with this binary would be at a too high frequency to be probed by \glspl*{pta}, but} the pulsar terms could still fall within their sensitivity band, thereby enabling the possibility of multi-band observations of \glspl*{smbhb} \cite{2013ApJ...764..187S}.
In this work, we extend beyond the case of binaries merging at present and show that \glspl*{pta} {are also sensitive to} past mergers of \gls*{smbhb}, due to the kiloparsec-scale distances between the monitored pulsars and Earth.
Hereafter, we refer to this new class of \gls*{pta} sources as \emph{zombie binaries}.
The paper is organized as follows.
In Section~\ref{sec:zombie-signal}, we first explain the principles of zombie binary detection with \glspl*{pta}.
We then estimate the likelihood of observing zombie binaries with high \gls*{snr} in \gls*{pta} datasets in Section~\ref{sec:zombie-number}.
This depends both on the \gls*{smbhb} population models considered, which we present in Section~\ref{sec:merger-density}, and on the \gls*{pta} sensitivity to zombie binary signals, which we estimate in Section~\ref{sec:det-efficiency}.
We then present our results for past, current, and future \gls*{pta} sensitivities in Section~\ref{sec:results}, before discussing the implications of our findings in Section~\ref{sec:discussion}.

In the following, $G$ is Newton's gravitational constant, $c$ is the speed of light in vacuum and $\log$ denotes the base-10 logarithm.

\section{\label{sec:zombie-signal}Zombie binary signals}

In \glspl*{pta}, the signal produced by a \gls*{gw} source consists of two contributions, usually referred to in the literature as the Earth term and the pulsar term \cite{1979ApJ...234.1100D}.
The arrival time of a radio pulse emitted by a pulsar $a$ is influenced by the metric perturbation both at the moment of its emission (at the pulsar) and at the moment of its reception (at Earth).
This is explicit in the expression of the induced timing residuals (see e.g. \cite{2010PhRvD..81j4008S}),
\begin{equation}
    \label{eq:timing-res-def}
        \RGW^{(a)}(t) = \sum_{A\in\{+,\times\}} F^A_a \int^{t} \dd{t^\prime}
        \left[ h_A\left(t^\prime\right) - h_A\left(t^\prime - \tau_a \right) \right],
\end{equation}
where $h_A$ denotes the plane-wave metric perturbation propagating along $\hat n$, and $F^A_a$ are the antenna pattern response functions of the pulsar for the $+$ and $\times$ polarization modes which, using Einstein summation on spatial indices $i, j$, can be expressed as
\begin{equation}
    \label{eq:antenna-pattern-function}
    F^A_a \equiv \frac{\hat{p}_a^i\hat{p}_a^je^A_{ij}(\hat n)}{2(1 + \hat n \cdot \hat p_a)}.
\end{equation}
We introduced the polarization tensors $e^A$ and  the unit position vector of pulsar $a$, $\hat{p}_a$.

In Eq.~\ref{eq:timing-res-def}, we see that while the Earth term of the timing residuals correspond to the integrated strain signal today, the pulsar term introduces a geometrical time delay,
\begin{equation}
    \label{eq:ET-PT-delay}
    \tau_a \equiv \left( 1 + \hat p_a \cdot \hat n\right) \Ta,
\end{equation}
where $\Ta \equiv D_a / c$ is the unperturbed time of flight to Earth of the radio photons\footnote{In this study, we do not consider any chromatic effect on the timing residuals and assume that all observations are done at a radio frequency of $1400$ MHz.} emitted by pulsar $a$.
For galactic pulsars monitored by \glspl*{pta}, the corresponding time delay is on the order of several thousand years and attains its maximum value of $2T_a$ when the \gls*{gw} source lies in the direction on the sky opposite to that of the pulsar.

In this work, we investigate the possibility to probe binaries that either merged before the \gls*{pta} experiment began or which are on the verge of merging.
For such systems, the corresponding Earth terms vanish or lie outside the \gls*{pta} band.
However, some pulsars can still exhibit non-zero pulsar terms for binaries that coalesced on timescales up to thousands of years in the past.
We therefore refer to this new class of \gls*{pta} sources as \emph{zombie binaries}.

At $0$-\gls*{pn} order, the orbital frequency evolution of \glspl*{smbhb} on circular orbit with dynamics driven by \gls*{gw} emission is, in the observer frame, given by \cite{1964PhRv..136.1224P}
\begin{equation}
    \label{eq:GW-freq-evol-obs}
    \frac{\dd \forb}{\dd t} = \frac{1}{2\pi}\frac{96}{5} \left( \frac{G \Mc}{c^3}\right)^{5/3} \left(2\pi\forb\right)^{11/3},
\end{equation}
where $\Mc = M_1^{3/5}M_2^{3/5} / (M_1 + M_2)^{1/5}$ is the observer-frame binary chirp mass.
Because Eq.~\ref{eq:GW-freq-evol-obs} strongly depends on the binary’s orbital frequency, $\forb$, a zombie binary that merged before the start of the \gls*{pta} experiment could still produce pulsar term signals within the \gls*{pta} frequency band.
Indeed, for a binary that coalesced at time $\tc$ in the observer frame, the orbital frequency of the binary as seen in the pulsar term of a given pulsar can be obtained by integrating Eq.~\ref{eq:GW-freq-evol-obs}, assuming that at $\tc$, we have $\forb = \forb^{\rm (ISCO)}$, the orbital frequency of the binary at its innermost stable circular orbit.
This yields,
\begin{equation}
    \label{eq:forbP-a}
    \forbP{a} = \forb^{\rm (ISCO)} \left[ 1 + \frac{\tau_a - \tauc}{\Tc\left(\forb^{\rm (ISCO)}\right)}\right]^{-3/8},
\end{equation}
where $\tauc \equiv t_0 - \tc$, is the time delay separating the starting date of the \gls*{pta} observations, $t_0$, from the binary merger.
We introduced the characteristic coalescence time of the binary from the orbital frequency $\forb$, $\Tc(\forb) \equiv 5 / 256 \left(G\Mc / c^3\right)^{-5/3} \left(2\pi \forb\right)^{-8/3}$.
We emphasize that we assume the binaries have been on circular orbits for the last few thousand years preceding their merger, with dynamics driven solely by \gls{gw} emission.
We therefore only consider the binaries \gls*{gw} radiation at $2\forb$ \cite{1963PhRv..131..435P}.
The effect of eccentricity, interaction with the binary disks as well as the impact of higher-order \gls*{pn} dynamics will be explored in future work.

\begin{figure}
    \centering
    \includegraphics[width=\linewidth]{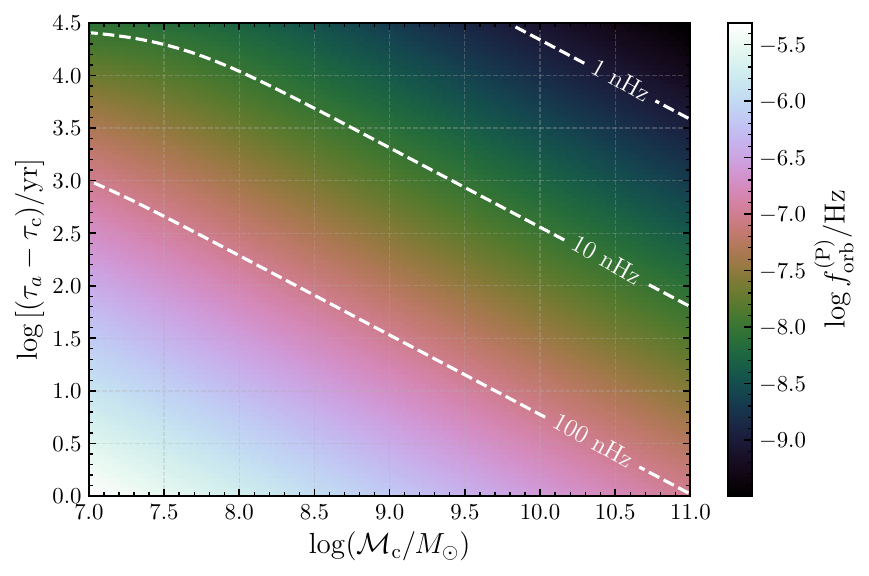}
    \caption{Initial orbital frequency of the binary in the pulsar term signal, expressed as a function of the observer-frame binary chirp mass and the time delay separating the pulsar term epoch from the binary merger, $\tau_a - \tauc$.
    We assume here that the binary mass ratio is $1$.}
    \label{fig:forbP-grid}
\end{figure}

Figure~\ref{fig:forbP-grid} illustrates how the zombie binary’s orbital frequency -- as observed in the pulsar term -- varies with both the binary chirp mass and the time delay separating the pulsar term epoch from the binary merger.
We see that for binaries with $\Mc \geq 10^9 \msun$, time delays of thousands of years are sufficient for the binary to have \gls*{gw} frequency well within the \gls*{pta} band, around $10$ nanohertz.
Therefore, zombie binaries that merged thousands of years before the start of any \gls*{pta} experiment could be observable in \gls*{pta} datasets as quasi-monochromatic sine waves whose frequency and amplitude are specific to each pulsar depending on their relative position to the binary and their distance from Earth.
However, a zombie binary will affect the timing residuals of a given pulsar $a$ only if $\tau_a > \tauc$.
{Otherwise, the pulsar term epochs observed by pulsar $a$ would correspond to the post-merger phase of the binary’s evolution, and thus will not be associated with any \gls*{gw} signal.}
Therefore, the further back in time the merger occurred, the less likely it is that pulsars in the \gls*{pta} will contribute to its detection.

\section{\label{sec:zombie-number}Occurrence of zombie binaries}

We now aim to quantify the probability of having {resolvable} zombie binaries in a given \gls*{pta} dataset.
{To assess the detectability of a zombie binary, we use the system’s optimal matched filter \gls*{snr} and rely on specific \gls*{snr} thresholds to estimate its detectability.}

In the following, we define zombie binaries as binaries whose Earth term orbital frequency at $t_0$, $\forbE$, exceeds a threshold frequency, $\fth$.
We note that this includes the case of binaries that merged before $t_0$, for which we consider that $\forbE = +\infty$.
In this paper, we use $\fth = 3 \times 10^{-7} \,\mathrm{Hz} \approx 1 / (2 \times 19 \,\mathrm{days})$, a frequency at the edge of the \gls*{pta} band.
We emphasize that increasing this threshold has a negligible impact on our results, since the number of binaries that merge during the \gls*{pta} observation (or shortly thereafter) is negligible compared to the number of binaries that have merged in the past thousands of years.

Since zombie binaries may already have coalesced at $t_0$, instead of distributing the \glspl*{smbhb} over their Earth term orbital frequency -- as is usually done in the \gls*{pta} literature \cite{2004ApJ...615...19E,2008MNRAS.390..192S,2019MNRAS.488..401C} -- we use their coalescence time as dynamical variable, through $\tauc$.
The distribution of merging \glspl*{smbhb} can be obtained from their comoving merger density, $n$, using
\begin{align}
    \label{eq:dNdtc-general}
    \frac{\dd^4 N}{\dd \xiI \dd \tauc} &= \frac{\dd^3 n}{\dd \xiI} \frac{\dd z}{\dd \taucr} \frac{\dd \Vc}{\dd z} \frac{\dd \taucr}{\dd \tauc} \\
    \label{eq:dNdtc-general-2}
    &= \frac{\dd^3 n}{\dd \xiI} \frac{4\pi c\dL^2}{(1+z)^2},
\end{align}
where we distribute the binaries over redshift, rest-frame chirp mass (expressed in base-10 logarithm of solar masses), and mass ratio, using the vector $\xiI = (z, \log \Mcr, q)$.
In Eq.~\ref{eq:dNdtc-general-2}, $\dL$ denotes the luminosity distance to a binary at redshift $z$ -- computed using the Planck 2018 cosmology \cite{2020A&A...641A...6P} -- which enters the expression for the differential comoving volume element, $\dd \Vc / \dd z$ \cite{1999astro.ph..5116H}.
Note that quantities evaluated in the binary rest frame are systematically indexed by the subscript $\mathrm{r}$.

Given our definition of zombie binaries, we can compute their average number across universe realizations by integrating Eq.~\ref{eq:dNdtc-general-2} over the corresponding $\tauc$ range. 
The first condition that a zombie binary must meet is that $\forbE \geq \fth$, since their associated Earth terms must be null or inaccessible to \glspl*{pta}.
Using Eq.~\ref{eq:forbP-a} with $\tau_a = 0$, this reads
\begin{equation}
    \label{eq:tauc-min}
    \tauc \geq \taucmin \equiv \Tc \left( \forbISCO\right) \left[ 1 - \left( \frac{\fth}{\forbISCO} \right)^{-8/3} \right].
\end{equation}
We note that, for highly massive binaries satisfying $\forbISCO \leq \fth$, this condition is replaced by $\tauc \geq 0$, to ensure that the binary merges before $t_0$.
In addition, a zombie binary can potentially be detected by a \gls*{pta} only if its inspiral can be observed at least by the pulsar term of the most distant pulsar of the array.
A zombie binary therefore must also verify $\tauc \leq 2 \Tmax$, where $\Tmax$ is the photons' time of flight associated with this most distant pulsar.

Consequently, the average number of zombie binary signals potentially present in a given \gls*{pta} dataset can be evaluated using
\begin{equation}
    \label{eq:avg-N-zombie}
    \langle N \rangle = \int \dd \xiI \frac{\dd^3 n}{\dd \xiI} \frac{4 \pi c \dL^2}{(1+z)^2} \left(2\Tmax - \taucmin \right),
\end{equation}
where we omitted the dependence of $\taucmin$ over $\xiI$.
Since a large fraction of these zombie binaries will correspond to very low \gls*{snr} values regardless of the \gls*{pta} considered, we instead compute the expectation value for the number of zombie binaries with a \gls*{snr} higher than a certain threshold $\SNRthresh$,
\begin{equation}
    \label{eq:N-zombie-SNR}
    \begin{aligned}
    \Nzbar \equiv \int \dd \xiI \frac{\dd^3 n}{\dd \xiI} &\frac{4 \pi c \dL^2}{(1+z)^2} \left(2\Tmax - \taucmin \right) \\
    \times & \:\mathrm{P} \left( \SNR \geq \SNRthresh \left| \xiI \right.\right),
    \end{aligned}
\end{equation}
introducing the detection efficiency of the \gls*{pta}, $\mathrm{P} \left( \SNR \geq \SNRthresh \left| \xiI \right.\right)$.
This function gives the probability that a zombie binary with intrinsic parameters, $\xiI$, yields a \gls*{snr} above $\SNRthresh$ for a given \gls*{pta} configuration.

{In the following, we consider $\SNRthresh \in \{3,5,10\}$. 
In principle, the \gls*{snr} threshold should be calibrated from the distribution of the detection statistic under the null hypothesis (i.e., noise-only datasets) for a prescribed false alarm probability \cite{kay1998fundamentals}.
However, estimating this distribution raises several methodological challenges that are beyond the scope of this study and will be addressed in future work.
We therefore consider a range of \gls*{snr} thresholds in order to assess the robustness of our results with respect to the threshold choice.}

The 3D integral of Eq.~\ref{eq:N-zombie-SNR} can be accurately approximated by the 2D integral over binary redshift and chirp mass.
Indeed, it is straightforward to verify that for the range of binary intrinsic parameters considered in this study\footnote{We note that for binaries with $\Mc \ll 10^8 \msun$, this approximation no longer holds. 
However, such low-mass systems are associated with a zero detection efficiency for the \glspl*{pta} considered in this work. 
Therefore, the region of the zombie parameter space with $\Mcr < 10^8 \msun$ does not contribute to the evaluation of Eq.~\ref{eq:N-zombie-SNR}.}, $2\Tmax - \taucmin \approx 2\Tmax$.
Furthermore, for comparable mass binaries, under the assumption of GW-driven dynamics at $0$-\gls*{pn} order, the binary mass ratio only enters in the zombies waveforms via $\forbISCO$ and thus have only a negligible effect on the detection efficiency.
We can therefore neglect the dependence over the binary mass ratio in Eq.~\ref{eq:N-zombie-SNR}.
For the remaining parameters, we consider $z\in[10^{-2}, 4]$ and $\log \Mcr / \msun \in [{7.5}, 11]$.

Since incorporating the \gls*{pta} detection efficiency in Eq.~\ref{eq:N-zombie-SNR} is equivalent to thinning the Poisson process \cite{daley2013introduction} governing the occurrence of merging \glspl*{smbhb}, the number of zombie binaries yielding a \gls*{snr} above $\SNRthresh$ follows a Poisson distribution of mean $\Nzbar$.
Once the latter is computed for a given comoving merger density model and a given \gls*{pta} sensitivity, the probability to have at least one zombie binary with $\SNR \geq \SNRthresh$ can be obtained using $\mathrm{P} \left( N_{\rm z} > 0\right) = 1 - e^{-\Nzbar}$.

\section{\label{sec:merger-density}Comoving merger density models}

In order to compute Eq.~\ref{eq:N-zombie-SNR}, one needs a model for the comoving \gls*{smbhb} merger density function, $n$.
We model it using the Schechter-like parametric model introduced in \cite{2016MNRAS.455L..72M},
\begin{equation}
    \label{eq:Prechter-based-model}
    \begin{aligned}
        \frac{\dd^2 n}{\dd z \dd \log \Mcr} = \:&\dot{n}_0 \left[ \left(\frac{\Mcr}{10^7 \msun}\right)^{-\alpha} e^{-\Mcr / \Mcstar} \right] \\
        &\times \left[ (1 + z)^\beta e^{-z/z_0} \right] \frac{\dd t_{\rm r}}{\dd z},
    \end{aligned}
\end{equation}
which depends on five parameters: the spectral indices $\alpha$ and $\beta$, the chirp mass and redshift exponential cutoffs $\Mcstar$ and $z_0$ and the comoving merger rate normalizing factor, $\dot{n}_0$.

In order to quantify the dependence of our results on the comoving merger density model, we follow \cite{2024PhRvD.109l3544S} and consider three different sets of parameters, given in Table~\ref{tab:pop-models}.
For each model, we adjust the value of $\dot{n}_0$ so that the \gls*{gw} background characteristic strain amplitude induced by the population is consistent with current \gls*{pta} observations.
Using the Phinney theorem \cite{2001astro.ph..8028P}, we set $h_{\rm c}(f=1/\mathrm{year}) = 3 \times 10^{-15}$. 

\begin{figure}
    \centering
    \includegraphics[width=\linewidth]{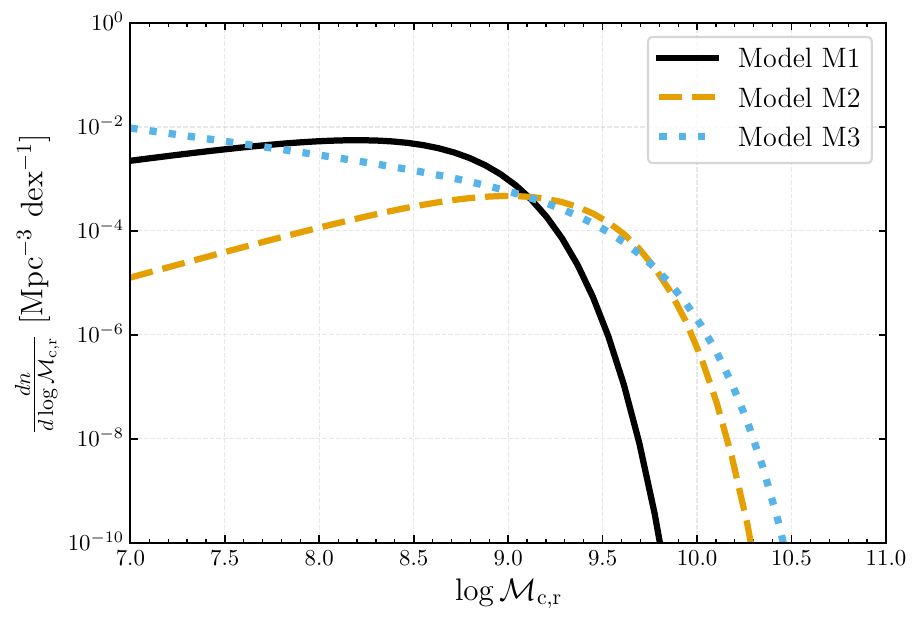}
    \caption{Comoving \gls*{smbhb} merger density integrated over redshift of the three population models considered in this work.}
    \label{fig:pop-merger-density}
\end{figure}

The comoving merger density associated with each model -- integrated over redshift -- are shown in Figure~\ref{fig:pop-merger-density}.
We note that our choice of models reflects the current uncertainty regarding the population of \glspl*{smbhb} likely to be responsible for the \gls*{pta} signal \cite{2024A&A...685A..94E,2023ApJ...952L..37A}.
For example, M1 includes a large number of relatively low-mass binaries, while M2 favors a smaller number of higher-mass systems.

\renewcommand{\arraystretch}{1.35}
\begin{table}[]
    \centering
    \begin{tabular}{|c|c|c|c|c|c|}
        \hline
         Name & $\alpha$ & $\log \Mcstar / \msun$ & $\beta$ & $z_0$ & $\dot{n}_0$ [Gyr$^{-1}$$\,$Mpc$^{-3}$] \\ \hline \hline
         M1 & $-0.5$ & $8.5$ & $1.5$ & $1.1$ & $1.83\times10^{-4}$ \\ \hline
         M2 & $-1$ & $9$ & $1$ & $0.5$ & $2.25\times10^{-6}$ \\ \hline
         M3 & $0.5$ & $9.3$ & $0.5$ & $1$ & $1.26\times10^{-3}$ \\ \hline
    \end{tabular}
    \caption{Parameters used for the three comoving merger density models employed in this work.}
    \label{tab:pop-models}
\end{table}

\section{\label{sec:det-efficiency}The zombie detection efficiency}

The second ingredient needed to evaluate $\langle \Nz \rangle$ is the detection efficiency which only depends on the \gls*{pta} configuration considered.
Since here, we neglect the influence of the binary mass ratio on $\mathrm{P} \left( \SNR \geq \SNRthresh \left| \xiI \right.\right)$, we numerically approximate it over a redshift and rest-frame chirp mass grid, fixing $q=1$.
For a given pair of intrinsic parameters, we generate ${10\,000}$ binaries drawing from uniform distribution their coalescence time $\tauc \in \left[\taucmin, 2\Tmax\right]$ and extrinsic parameters: sky location, polarization angle, inclination angle and orbital phase at coalescence.
We then compute the \gls*{snr} of each of these zombie binaries and approximate $\mathrm{P} \left( \SNR \geq \SNRthresh \left| \xiI \right.\right)$ taking the fraction of binaries satisfying $\SNR \geq \SNRthresh$.

To compute the zombies' optimal matched filter \glspl*{snr}, we sum in quadrature the squared \gls*{snr} contributions from each pulsar of the \gls*{pta}:
\begin{equation}
    \SNR = \sqrt{\sum_{a=1}^{\Npsr} \vec{R}_{\rm z}^{(a) \mathrm{T}} \textbf{C}_{a}^{-1} \vec{R}^{(a)}_{\rm z}},
\end{equation}
where $\vec{R}^{(a)}_{\rm z}$ are the timing residuals induced by the zombie binary in pulsar $a$ and $\textbf{C}_{a}$ is the noise covariance matrix of pulsar $a$.
We emphasize that, to confidently detect a zombie binary, it is not enough to have a high \gls*{snr}: the signal must also be observed in the data of multiple pulsars\footnote{{We recall that if $\tau_a < \tauc$ for pulsar $a$, $\vec{R}^{(a)}_{\rm z} = \vec{0}$, such that pulsar $a$ is not contributing to the zombie \gls*{snr}.}}. 
Otherwise, the possibility that the signal arises purely from intrinsic noise cannot be ruled out.
{Therefore, in the following, we do not consider a zombie binary to be detectable if only a single pulsar contributes to its \gls*{snr}, even if the latter exceeds the \gls*{snr} threshold.}

We evaluate the noise covariance matrices of pulsars following \cite{2014PhRvD..90j4012V} and using the \textsc{enterprise} Python software \cite{enterprise}.
We incorporate both the radiometer measurement noise \cite{2018CQGra..35m3001V}  -- quantified by its root-mean-square error, $\sigma_{\rm MN}$ -- and the red noise process associated with the astrophysical \gls*{gw} background.
We model the latter using a power-law power spectral density model with a spectral index of $13/3$ -- typical for a population of circular binaries driven by \gls*{gw} emission \cite{2001astro.ph..8028P} -- and a corresponding characteristic strain amplitude at $1/$year of $3\times10^{-15}$.

We also include the effect of timing model marginalization (see \cite{2019PhRvD.100j4028H} and references therein) in each pulsar covariance matrix.
In particular, we incorporate the effect of marginalizing on the pulsars' spin frequency and derivative as well as on the pulsars' sky position, which significantly reduces the sensitivity of the array at \gls*{gw} frequencies below its fundamental frequency, $1/\Tobs$, and around the $1/$year frequency, respectively.

We note that the \gls*{pta} detection efficiencies that we derive neglect the presence of potential intrinsic red noise components -- such as pulsar spin noise or dispersion measure variation noise -- and do not take into account the spatially-correlated nature of the \gls*{gw} background noise.
However, these assumptions should not have strong impact on our results, since these red noise components typically have a sub-dominant contribution to \gls*{pta} sensitivities at the frequencies where zombie binaries are found (see Section~\ref{sec:results}) \cite{2015CQGra..32e5004M,2019PhRvD.100j4028H}.
Instead, the timing measurement precision of the \gls*{pta} is expected to have a strong impact on the detectability of zombie binaries.

\begin{table}[]
    \centering
    \begin{tabular}{|c|c|c|c|c|c|}
    \hline
        PTA & $\Tobs$ [yr] & $\Npsr$ & $\Delta t$ [days] & $\sigma_{\rm MN}$ & $\Tmax$ [kyr] \\ \hline \hline
        EPTA & 10 & 25 & 5 & 1 $\mu$s & 12.8 \\ \hline
        IPTA & 10 & 130 & 5 & 1 $\mu$s & 12.8 \\ \hline
        SKA & 10 & 130 & 7 & 50 ns & 20.4 \\ \hline
    \end{tabular}
    \caption{\glspl*{pta} configurations considered in this work.
    The name, observing time, number of pulsars, observing cadence, root-mean-square timing measurement error, and the maximum photon travel time are shown for each array.}
    \label{tab:PTA-configs}
\end{table}

\section{\label{sec:results}Results}

\begin{figure*}
    \centering
    \includegraphics[width=\linewidth]{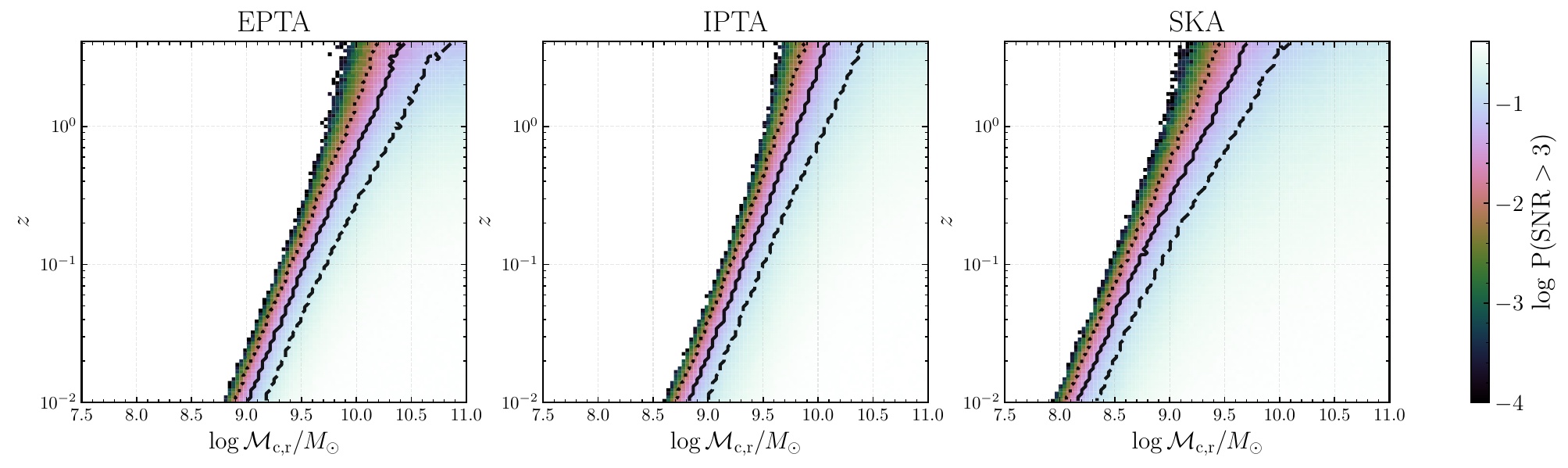}
    \caption{Detection efficiency {for $ \SNRthresh = 3$} of zombie binaries as a function of redshift and rest-frame chirp mass, for the three \glspl*{pta} configurations considered. 
    See Table~\ref{tab:PTA-configs} for details on the settings used for each array.
    Because our estimate of the detection efficiency is based on ${10\,000}$ simulated binaries, our sensitivity is restricted to probabilities above $10^{-4}$.
    {We overlay, using dotted, solid, and dashed lines, the contours representing $ \mathrm{P}(\mathrm{SNR} > \SNRthresh) \geq 1\%$ with $\SNRthresh = 3$, $5$ and $10$, respectively.}}
    \label{fig:log-efficiency}
\end{figure*}

In order to quantify the capacity of \glspl*{pta} to probe zombie binaries, we consider representative configurations for past, current and future \gls*{pta} datasets, which are summarized in Table~\ref{tab:PTA-configs}.
They aim to represent respectively the sensitivity of the European PTA (EPTA) second data release (referred to as DR2new in \cite{2023A&A...678A..48E}), the upcoming International PTA (IPTA) third data release (DR3) and a potential future sensitivity of the Square Kilometer Array (SKA) observatory \cite{2025OJAp....854243S}.
For the IPTA DR3-like sensitivity, we consider $130$ pulsars present in the previously published collaboration datasets, fixing their observing time to $10$ years, with an observing cadence of $5$ days.
For the SKA configuration, we use $124$ pulsars of our IPTA-like dataset with declination angle below $45^\circ$ \cite{2015aska.confE..37J} and added randomly $6$ pulsars from \cite{cornell_parallax}.
For pulsars whose distances to Earth could not be estimated, we randomly assign them a distance from $0.5$ to $2.5$ kiloparsecs using uniform weighting. 

In Figure~\ref{fig:log-efficiency}, we show the detection efficiency for zombie binaries, evaluated for each \gls*{pta} configuration{, under the most optimistic detection threshold, $\SNRthresh = 3$.}
We find that the EPTA DR2new-like dataset can only probe massive systems with $\Mcr \approx 10^9 \msun$ at redshift below $0.03$.
However, such \glspl*{pta} could probe the very massive -- but rare -- mergers with $\Mcr \approx 10^{10} \msun$ up to $z \approx 1$.
Since the \glspl*{snr} of zombie binaries scale as $\sqrt{\Npsr}$, the situation improves only slightly for our IPTA DR3 setting.
However, the true IPTA DR3 dataset will be much more heterogeneous with some pulsars having an observing time longer than $10$ years with a better cadence, as well as lower timing precision \cite{2024ApJ...966..105A}.
This could significantly enhance the detection efficiency of this dataset and we therefore regard our IPTA sensitivity as a conservative baseline.
Going to the SKA configuration, we find that zombie binaries of a few billion solar masses (or above) {are associated with high detection efficiency} up to redshift $1$.
Less massive mergers with $\Mcr < 10^9\msun$  {only have a reasonable chance to exceed the detection threshold} providing they happened at very low redshift, typically at $z \lesssim 0.1$.

{In order to quantify the dependence of the zombie detection efficiency on the \gls*{snr} threshold considered, we overlay in Figure~\ref{fig:log-efficiency} the contours corresponding to $\mathrm{P}(\mathrm{SNR} > \SNRthresh) \geq 1\%$, with $\SNRthresh = 3$, $5$, and $10$, shown as dotted, solid, and dashed lines, respectively.
At low redshift ($z \lesssim 0.1$), and low chirp masses ($\log \Mcr \lesssim 9.5$), the shifts of the contours toward higher chirp masses simply reflect the scaling of the zombie \gls*{snr} with the latter\footnote{{Indeed, combining Eqs.~\ref{eq:timing-res-def} and \ref{eq:forbP-a}, it is straightforward to show that the average zombie binary timing residual amplitude scales as $\Mc^{15/8}$. Therefore, in Eq.~\ref{eq:N-zombie-SNR}, if the noise contribution is assumed to be constant across binaries, we expect, e.g., a factor of $(10 / 3)^{8/15} \approx 1.9$ between the dotted and dashed contours in Figure~\ref{fig:log-efficiency}, which is indeed recovered.}}.
However, at higher redshift and for higher-mass systems, this simple scaling no longer holds. This is due to the red astrophysical \gls*{gw} background noise, which starts to dominate \glspl*{pta}' sensitivities at low frequency. 
As opposed to the white measurement noise, this background noise more strongly limits the \gls*{snr} of zombie binaries as they reach lower orbital frequencies. 
Consequently, since higher-mass systems evolve more rapidly and thus reach lower pulsar-term \gls*{gw} frequencies, the scaling of the zombie \gls*{snr} deviates from the low-mass limit.
A similar reasoning applies to high-redshift systems, which are also associated with lower redshifted \gls*{gw} frequencies.}

\begin{table*}[t!]
\small
\setlength{\tabcolsep}{2.5pt}
\renewcommand{\arraystretch}{0.95}
\centering
\resizebox{\textwidth}{!}{%
\begin{tabular}{c|cccccc|cccccc|cccccc}
    \toprule
      & \multicolumn{6}{c}{M1} & \multicolumn{6}{c}{M2} & \multicolumn{6}{c}{M3} \\
     & \multicolumn{2}{c}{$\mathrm{SNR}>3$} & \multicolumn{2}{c}{$\mathrm{SNR}>5$} & \multicolumn{2}{c}{$\mathrm{SNR}>10$} & \multicolumn{2}{c}{$\mathrm{SNR}>3$} & \multicolumn{2}{c}{$\mathrm{SNR}>5$} & \multicolumn{2}{c}{$\mathrm{SNR}>10$} & \multicolumn{2}{c}{$\mathrm{SNR}>3$} & \multicolumn{2}{c}{$\mathrm{SNR}>5$} & \multicolumn{2}{c}{$\mathrm{SNR}>10$} \\
     & $p_{N_{\mathrm{z}}}$ & $\langle N_{\rm z} \rangle$ & $p_{N_{\mathrm{z}}}$ & $\langle N_{\rm z} \rangle$ & $p_{N_{\mathrm{z}}}$ & $\langle N_{\rm z} \rangle$ & $p_{N_{\mathrm{z}}}$ & $\langle N_{\rm z} \rangle$ & $p_{N_{\mathrm{z}}}$ & $\langle N_{\rm z} \rangle$ & $p_{N_{\mathrm{z}}}$ & $\langle N_{\rm z} \rangle$ & $p_{N_{\mathrm{z}}}$ & $\langle N_{\rm z} \rangle$ & $p_{N_{\mathrm{z}}}$ & $\langle N_{\rm z} \rangle$ & $p_{N_{\mathrm{z}}}$ & $\langle N_{\rm z} \rangle$ \\
    \midrule
    EPTA & $1.9\!\times\! 10^{-4}$ & 0.00 & $3.2\!\times\! 10^{-5}$ & 0.00 & $2.4\!\times\! 10^{-6}$ & 0.00 & 0.03 & 0.03 & $7.4\!\times\! 10^{-3}$ & 0.01 & $9.4\!\times\! 10^{-4}$ & 0.00 & 0.03 & 0.03 & $7.4\!\times\! 10^{-3}$ & 0.01 & $9.6\!\times\! 10^{-4}$ & 0.00 \\
    IPTA & $2.8\!\times\! 10^{-3}$ & 0.00 & $5.1\!\times\! 10^{-4}$ & 0.00 & $4.9\!\times\! 10^{-5}$ & 0.00 & 0.23 & 0.27 & 0.07 & 0.07 & 0.01 & 0.01 & 0.22 & 0.25 & 0.07 & 0.07 & 0.01 & 0.01 \\
    SKA & 0.90 & 2.31 & 0.41 & 0.53 & 0.06 & 0.07 & 1.00 & 6.15 & 0.94 & 2.75 & 0.51 & 0.71 & 1.00 & 5.73 & 0.90 & 2.30 & 0.42 & 0.54 \\
    \bottomrule
\end{tabular}}
\caption{{Occurrence probabilities, $p_{N_{\mathrm{z}}}$, and associated expected number of zombie binaries above the detection threshold, $\langle N_{\rm z} \rangle$, for each population model and \gls*{snr} detection threshold. 
We recall that a zombie binary must be observed by at least two pulsars of the array to meet the detection criteria.}}
\label{tab:pNz}
\end{table*}

{Using the numerical detection efficiencies associated with the different \gls*{snr} thresholds}, we then evaluate Eq.~\ref{eq:N-zombie-SNR} for each of the three \glspl*{smbhb} population models presented in Section~\ref{sec:merger-density}.
Our results are summarized in {Table~\ref{tab:pNz}, where we show the probability to have at least one zombie binary with \gls*{snr} above $\SNRthresh$ for each \gls*{pta}-population model pair}.
While EPTA-like datasets are not likely to be able to detect any zombie binary -- $\mathrm{P} \left( N_{\rm z} > 0\right) \leq 3\%$ {for all models, with $\SNRthresh=3$} -- the situation strongly improves for IPTA-like and SKA-like datasets.
Although the IPTA configuration has a very small chance to {have bright zombie binaries in its data} under M1 -- due to its limited detection efficiency for systems with $\Mcr < 10^9 \msun$ -- the probability increases to $23\%$ and $22\%$ for M2 and M3, respectively, {considering $\SNRthresh=3$.
However, these probabilities both drops to $1\%$ under the most conservative threshold, $\SNRthresh=10$.}
Notably, for the SKA-like \gls*{pta}, $\mathrm{P} \left( N_{\rm z} > 0 \, \mid \SNRthresh=3\right) \geq 90\%$, for all three population models.
{We find that such dataset should have, on average, $2.3$, $6.1$, and $5.7$ zombie binaries with SNR larger than 3}, considering the M1, M2 and M3 models, respectively.
Even in a far more pessimistic scenario where $\sigma_{\rm MN} = 300$ ns for all SKA pulsars, $\langle \Nz \rangle$ remains around $1$ for both the M2 and M3 models -- although it drops to $0.06$ for M1.
{Remarkably, even under the detection threshold $\SNRthresh=10$, the occurrence probability remains fairly high for the M2 ($51\%$) and M3 ($42\%$) models.
It therefore appears likely that, if the spatially-correlated signal observed by the \gls*{pta} collaborations truly arises from a population of \glspl*{smbhb}, the forthcoming SKA dataset will include \gls*{gw} signatures from at least one past binary merger, exhibiting an optimal matched filter \gls*{snr} in the range $[3, 10]$.}

\begin{figure*}[t!]
    \centering
    \includegraphics[width=\linewidth]{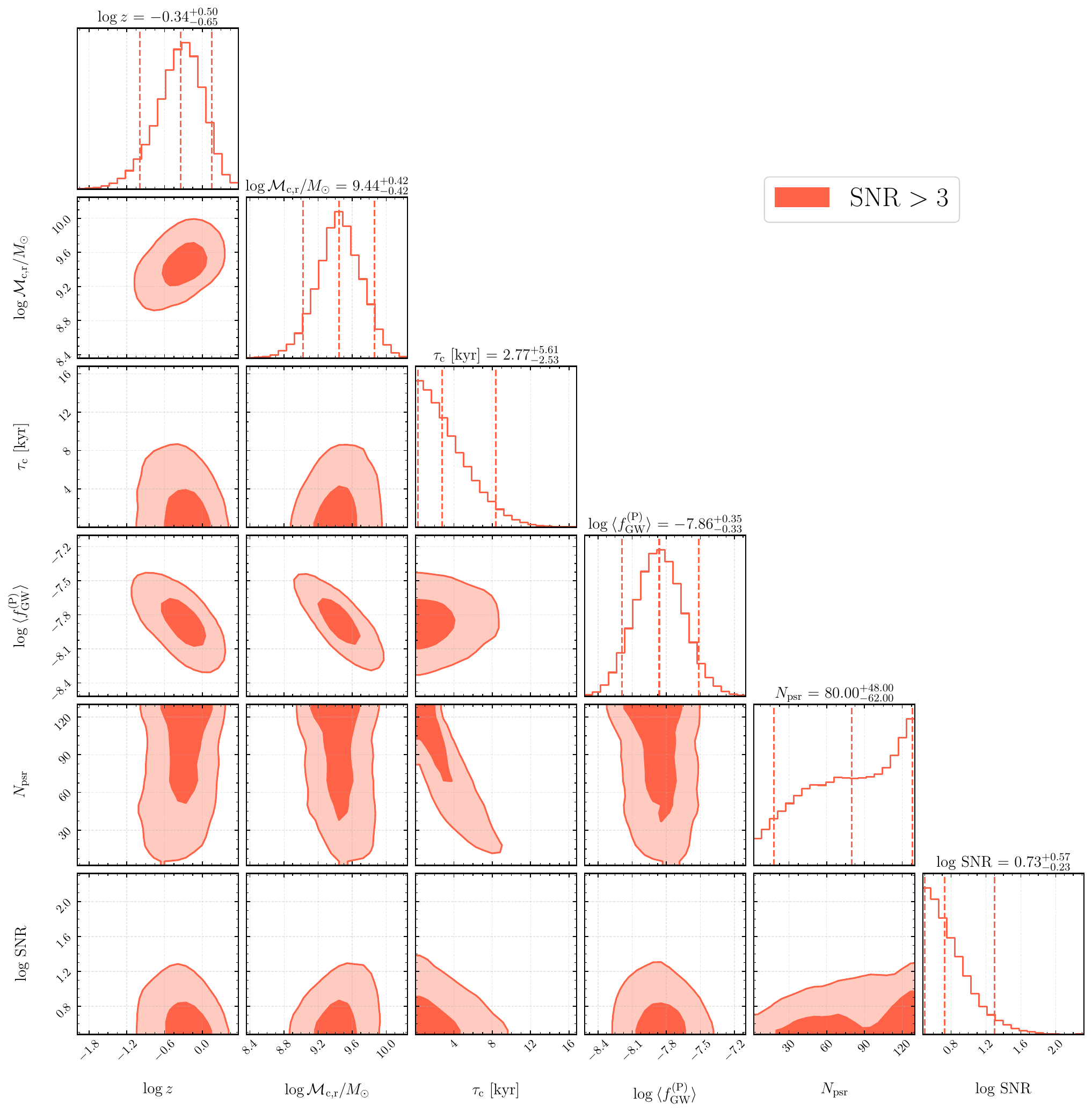}
    \caption{2D-distribution of the properties of zombie binaries with a \gls*{snr} greater than $3$ in the SKA configuration, assuming the M3 population model.
    In addition to the intrinsic properties of the merger, we report the number of pulsars $\Npsr$ that actually observe the binary before its coalescence -- and therefore contribute to its potential detection -- and the average \gls*{gw} frequency of their pulsar terms, $\langle \fgw^{\rm (P)} \rangle$.}
    \label{fig:zombie-props}
\end{figure*}

To evaluate the properties of the zombie binaries that the SKA observatory {might be able to detect} in the future, we sample the distribution, $p(\xiI)$, proportional to the integrand of Eq.~\ref{eq:N-zombie-SNR}.
Each zombie binary is then assigned a coalescence time and a set of extrinsic parameters -- we recall that we assume uniform distributions for the latter.
We then select only systems with a \gls*{snr} higher than $3$ for the SKA configuration.
We find that the binary population model considered does not significantly impact the distribution of the zombie binaries' parameters.
Therefore, in Figure~\ref{fig:zombie-props}, we only show the distribution obtained using the M3 model.

We find that the bright zombie binaries are massive systems with $\Mcr > 10^9 \msun$, located at low redshift: $95\%$ of them are found at $z<1.5$, under M3.
These high masses are required in order to have a significant frequency evolution in the thousands of years separating the pulsar terms' epochs from the binary coalescence.
Despite the \textit{a priori} uniform distribution on $\tauc$, we find that the bright zombie binaries are typically merging close to the \gls*{pta} starting date: less than $5\%$ of these systems merge more than $8.4$ thousand years before $t_0$.
This behavior is expected: as shown in the corresponding cells of Figure~\ref{fig:zombie-props}, the smaller $\tauc$ is, the more pulsars can observe the binary before its merger, and thus the higher its \gls*{snr} becomes.
{As a result, as illustrated in the $\Npsr$ column of Figure~\ref{fig:zombie-props}, the number of pulsars contributing to the \gls*{snr} of zombie binaries generally reaches several dozen for the SKA configuration.}
This could allow their identification through the sky-dependent frequency signature of their pulsar terms: a distinctive feature of these systems.
{In Appendix~\ref{app:snr2-cumul}, we quantify more precisely the number of contributing pulsars by determining how many are required to achieve $90\%$ of the \gls*{snr}$^2$ threshold.}

Finally, since the coalescence of these binaries of a few billion solar masses typically occurs thousands of years after the pulsar terms' epochs, {we expect} the \gls*{gw} frequencies associated to the latter {to be} found around $10-20$ nanohertz (as predicted in Figure~\ref{fig:forbP-grid}).
{This is confirmed in Figure~\ref{fig:zombie-props}, which displays the distribution of the mean \gls*{gw} frequency, $\langle \fgw^{\rm (P)} \rangle$, associated with the pulsar terms of the zombie binaries. 
To obtain this mean frequency for a given system, we consider only those pulsars whose pulsar terms are sensitive to the binary inspiral and thus contribute to the \gls*{snr} calculation.
We find, under M3, that $90\%$ of the bright zombie binaries verify $\log \,\langle \fgw^{\rm (P)} \rangle \in [-8.2, -7.5]$\footnote{{We note that under the M1 population model, which admits a lower chirp mass cutoff, this interval shifts towards higher frequencies, $\log \langle \fgw^{\rm (P)} \rangle \in [-7.85, -7.34]$, due to the slower dynamical evolution of lower-mass systems.}}.}
Incidentally, this represents an optimal range that lies between the reduced sensitivity of \glspl*{pta} at higher frequencies (mainly caused by the timing precision, observing cadence and the response of timing residuals to \glspl*{gw}) and the low-frequency regime where red noise dominates (arising from both the \gls*{gw} background(s) and intrinsic red noise processes).
{This frequency range where bright zombie binaries are most likely found could thus mitigate false alarm detections caused by particular red noise realizations.
A dedicated detectability analysis based on simulated \gls*{pta} datasets is required to investigate this hypothesis further.}

\section{\label{sec:discussion}Discussion}

In this work, we demonstrated that \glspl*{pta} have the capacity of probing the mergers of \glspl*{smbhb} that occurred before the start of the experiment.
This new class of \gls*{pta} sources -- which we call zombie binaries -- are accessible thanks to the time delays of several thousands of years that separate the so-called pulsar term from the Earth term in the timing residuals induced by a \gls*{gw} source.

Considering the case of circular binaries driven by \gls*{gw} emission, we estimated the number of zombie binaries with a \gls*{snr} greater than 3 that past (EPTA DR2), current (IPTA DR3) and future (SKA) \gls*{pta} datasets are likely to be sensitive to.
In order to capture the current uncertainty on the merger rate of \glspl*{smbhb}, we considered three analytic comoving merger density models -- based on Schechter-like functions -- compatible with current \gls*{pta} observations.
While EPTA DR2-like datasets lack sensitivity to yield a significant probability of detecting zombie binaries, we find that, for \glspl*{smbhb} models favoring large mass systems (with $\Mcr > 10^9 \msun$), the IPTA DR3 dataset has a $23\%$ probability of containing at least one zombie binary with $\SNR > 3$.
For the SKA sensitivity considered here, we find that, for all population models, this probability is greater than $90\%$.
We predict that a few bright zombie binaries {-- with \gls*{snr} in the range $[3,10]$ --} are likely to be {present} in future SKA datasets if the currently observed \gls*{pta} signal is sourced by a population of \glspl*{smbhb}.
These massive mergers ($\Mcr > 10^9 \msun$) associated with pulsar term signals' frequency on the order of $10$ nanohertz, will typically be probed at low redshift ($z \lesssim 1$).
However, we note that these findings rely on the assumption that \glspl*{smbhb} are on circular orbits.
If the signal observed by \glspl*{pta} originates from a population of binaries that are on very eccentric orbits when entering the \gls*{pta} band -- as recent studies suggest \cite{2024A&A...691A.212R,2025A&A...703A..86F} -- then zombie binaries may still have significant eccentricities.
Because the orbital evolution of eccentric binaries is faster, zombie binaries could be found at lower orbital frequencies than in the circular case.
This should be explored in future work.

We also note that the \gls*{snr} calculations used in this work do not account for our current uncertainty about the pulsar distances and therefore could be considered optimistic.
However, precise distance measurements using very long baseline interferometry are becoming increasingly common for millisecond pulsars \cite{2019ApJ...875..100D}.
In addition, the significant improvement in timing precision allowed by the SKA is expected to enable pulsar distance measurements with less than $5\%$ error for the pulsars' distances considered in this work \cite{2011A&A...528A.108S}.
Achieving a distance accuracy for pulsars on the order of hundreds of light-years will allow precise determination of the frequencies of the pulsar terms. 
This will help the coherent reconstruction of zombie binary signals across multiple pulsars, which is required for a future robust detection.

{It is also important to recall that the optimal matched filter \gls*{snr} considered in this work does not capture the complexity of realistic pulsar noise models, which include intrinsic noise components along with different potential common correlated signals.
In addition, the SKA will most likely be capable of detecting individual \glspl*{smbhb} \cite{2009MNRAS.394.2255S,2026A&A...706A.115T} whose signals could partially overlap with those of zombie binaries.
Although the latter only significantly affect a subset of pulsars and yield different \gls*{gw} frequencies in each pulsar of the array, the simultaneous search for standard continuous wave signals from \glspl*{smbhb} could also affect the detectability of zombie binaries.
The assessment of zombie binaries' detectability in realistic simulated \gls*{pta} datasets is deferred to future work.}

Such indirect detection of past binary mergers would open a remarkable window on the study of the environment and dynamics of \glspl*{smbhb} during the last thousands of years before coalescence.
However, it will also represent a challenge for detecting their signals, since the waveform associated with each pulsar of the array depends on the orbital frequency evolution model considered.
In any case, the search for zombie binaries will -- and could already -- put new constraints on the high-mass end of \glspl*{smbhb} merger rate models.

\section*{Data Availability}

The data and scripts used to produce the findings of this work are openly available at \href{https://github.com/hquelquejay/zombie-binaries-pta}{https://github.com/hquelquejay/zombie-binaries-pta}.

\begin{acknowledgments}
This work was supported by the European Union’s H2020 ERC Advanced Grant ``PINGU'' (Grant Agreement No. 101142079).
HQL is grateful to A. Sesana for helpful discussions.
HQL also thanks A. Sesana, S. Babak and M. Falxa for their comments on this manuscript.
HQL thanks the anonymous referee for constructive comments that greatly improved the manuscript.
Numerical computations were partly performed on the CC-IN2P3 (CNRS), in France.

\textit{Softwares:} \textsc{astropy} \cite{astropy:2022}, \textsc{corner} \cite{corner}, \textsc{jax} \cite{jax2018github}, \textsc{numpy} \cite{harris2020array}, \textsc{scipy} \cite{2020SciPy-NMeth}.
\end{acknowledgments}

\medskip

\appendix

\section{\label{app:snr2-cumul}NUMBER OF CONTRIBUTING PULSARS TO THE ZOMBIE SNRs}

\begin{figure*}
    \centering
    \includegraphics[width=\linewidth]{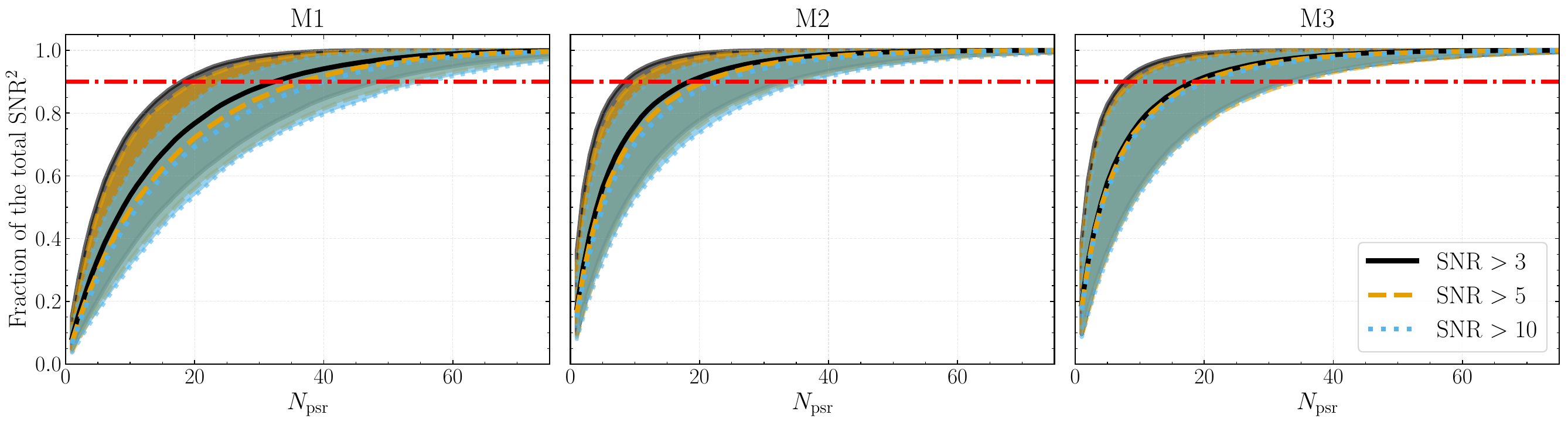}
    \caption{Distribution of the cumulative fraction of the total $\SNR^2$ for bright zombie binaries -- as seen by the SKA configuration -- shown for different population models and various total \gls*{snr} range.
    Note that we limit the visualization to $\Npsr \le  75$, although the SKA configuration includes $130$ pulsars.
    The medians and corresponding $68\%$ confidence intervals are represented by lines and shaded regions, respectively.
    For the M2 and M3 models -- displayed in the middle and right panels, respectively -- we find that, for half of the bright systems, the contributions from roughly $20$ pulsars must be combined to recover at least $90\%$ of the total $\SNR^2$ of the source.}
    \label{fig-app:snr2-cumul}
\end{figure*}

In this section, we aim to determine how many pulsars typically make a significant contribution to the total \gls*{snr} of bright zombie binaries.
To this end, we use the bright zombie binary samples associated with the different population models, as shown in Figure~\ref{fig:zombie-props} for M3.
For each system, we calculate the individual optimal matched filter $\SNR^2$ for every pulsar in the array.
We then sort these individual contributions in descending order and evaluate the fraction of the total squared \gls*{snr} as a function of the number of pulsar contributions included.
Combining these contributions from each bright zombie binary, we estimate the distribution of this cumulative $\SNR^2$ fraction function. 
The results are shown for the SKA configuration in Figure~\ref{fig-app:snr2-cumul}, where we display both the median and the $68\%$ confidence intervals of the obtained distributions.
The M2 and M3 models yield very similar results. In most cases, the number of pulsars required to reach $90\%$ of the total squared \gls*{snr} ranges from $6$ to $35$, with an almost negligible dependence on the total \gls*{snr} of the source.
Conversely, bright zombie binaries in the M1 model require a larger number of pulsar contributions to attain such a fraction of the total signal power. 
This is due to the fact that bright zombie binaries in M1 have lower chirp masses than the bright systems found in M2 and M3, and therefore undergo a slower dynamical evolution. 
Consequently, the \gls*{gw} frequencies of their associated pulsar terms are higher and therefore yield smaller timing residual amplitudes. 
These binaries thus need to be observed by a larger number of pulsars in order to reach a significant squared \gls*{snr}.

\bibliography{apssamp}

\end{document}